\documentclass[pra,showpacs,preprintnumbers,amsmath,aps,twocolumn]{revtex4}
\usepackage{amssymb}
\usepackage{txfonts}
\usepackage{amssymb}
\usepackage{amssymb}
\usepackage{graphicx}
\usepackage{color}
\usepackage{amsmath}
\usepackage{dcolumn}
\usepackage{bm}

\newcommand{\bei}{\begin{itemize}}
\newcommand{\eei}{\end{itemize}}
\newcommand{\bee}{\begin{enumerate}}
\newcommand{\eee}{\end{enumerate}}

\begin{document}
\title{Adiabatic Condition and Quantum Geometric Potential}
\author{Jian-da Wu$^{1,4}$}
\email{jdwu@mail.ustc.edu.cn}
\author{Mei-sheng Zhao$^1$, Jian-lan Chen$^3$}
\author{ Yong-de Zhang$^{2,1}$}

\affiliation{$^1$Hefei National Laboratory for Physical Sciences at
Microscale and Department of Modern Physics, University of Science
and Technology of China, Hefei 230026, People's Republic of China
\\$^2$CCAST (World Laboratory), P.O.Box 8730, Beijing 100080,
People's Republic of China
\\$^3$School of physics and material science, Anhui University, Hefei 230039, People's Republic of China
\\$^4$Department of Physics $\&$
Astronomy, Rice University, Houston, Texas 77005, USA}

\date{\today}

\begin{abstract}
In this paper, we present a $U(1)$-invariant expansion theory of the
adiabatic process. As its application, we propose and discuss new
sufficient adiabatic approximation conditions. In the new
conditions, we find a new invariant quantity referred as quantum
geometric potential (QGP) contained in all time-dependent processes.
Furthermore, we also give detailed discussion and analysis on the
properties and effects of QGP.

\end{abstract}


\maketitle

Since the establishment of the quantum adiabatic theorem
\cite{Ehrenfest,Born,Schwinger,Kato} in 1923, many fundamental
results have been obtained, such as Landau-Zener transition
\cite{phs.z.sowjetunion}, the Gell-Mann-Low theorem
\cite{Gell-Mann}, Berry phase \cite{Berry} and holonomy
\cite{Simon}. Also the adiabatic processes find their applications
in the quantum control and quantum computation
\cite{Oreg,J.A.Jone,Farhi,Zheng}. Recently the common-used
quantitative adiabatic condition \cite{Schiff,Messiah,Landau} has
been found not able to guarantee the validity of the adiabatic
approximation \cite{Marzlin,Tong1}. Consequently various new
conditions are conjectured and a series of confusions and debates
arise. For example, it was argued \cite{Wu} that the traditional
adiabatic condition did not have any problem at all and that the
invalidation of the condition did not mean the invalidation of
adiabatic theorem \cite{Duki}. Some new conditions proposed in
\cite{Ye,Tong2} but too rigorous to be used conveniently. Although
[22] also adopted the adiabatic perturbation expansion but did not
give out proper condition because the basis in \cite{MacKenzie} can
not show certain geometric properties in the adiabatic process.
\cite{Vertesi} pointed out the limitation of traditional condition
but also did not give out a proper condition. To solve the problem
of insufficiency of traditional adiabatic condition in
\cite{Marzlin,Tong1} and clarify the subsequent confusions, in this
paper, we present two new sufficient conditions in which the
properties and effects of a new invariant quantity are detailedly
discussed.

Let us consider a quantum system governed by a time dependent
Hamiltonian $H(t)$ and the initial state of the system is an
eigenstate $|m,0\rangle$ of $H(0)$ with eigenvalue $E_m (0)$, where
$m$ denotes the initial value of dimensionless quantum number set.
By introducing a dimensionless time parameter $\tau = E_m \left( 0
\right)t/\hbar $ and a dimensionless Hamiltonian $h(\tau ) = H(\tau
)/E_m \left( 0 \right)$,  the time dependent Schr\"odinger equation
reads

\begin{equation}\label{g1}
 i\frac{\partial |\Phi_m(\tau)\rangle}{\partial \tau }
 = h(\tau )|\Phi_m(\tau)\rangle,\quad
 \left|\Phi_m( 0)\right\rangle = \left| {m,0} \right\rangle.
\end{equation}
The exact solution $\left| {\Phi_m \left( \tau \right)}
\right\rangle$ to Eq.(\ref{g1}) is referred to as the system's
$dynamic$ $evolution$ $orbit$ in the Hilbert space.

Furthermore, by considering $\tau$ as a fixed parameter, we can
always solve the following quasi-stationary equation
\begin{equation}
h\left( \tau  \right)\left| {\varphi _n \left( \tau  \right)}
\right\rangle  = e_n \left( \tau  \right)\left| {\varphi _n \left(
\tau  \right)} \right\rangle. \label{g2}
\end{equation}
And the eigenstate $|\varphi_n(\tau)\rangle$ with the corresponding
initial state $\left| {n,0} \right\rangle$ is referred to as the
$adiabatic$ $orbit$ of the system.

For convenience, we denote $\gamma _{nm}  \equiv i\left\langle
{{\varphi _n (\tau )}}
 \mathrel{\left | {\vphantom {{\varphi _n (\tau )} {\dot \varphi _m (\tau )}}}
 \right. \kern-\nulldelimiterspace}
 {{\dot \varphi _m (\tau )}} \right\rangle$ and the dot here and below
expresses the derivative with respect to time. Apparently, an
adiabatic orbit multiplied by an arbitrary time-dependent phase
factor still describes the same adiabatic orbit. It is not difficult
to see that the following adiabatic orbit
\begin{equation}
\left| {\Phi _m^{adia} (\tau)} \right\rangle  = \exp\left\{ -
i\int_0^\tau  [e_m (\lambda ) -\gamma _{mm}(\lambda)]d\lambda
\right\}| {\varphi _m (\tau )}\rangle \label{a4}
\end{equation}
is invariant under the following $U(1)$ transformation
\begin{equation}\label{gauge}
\left| {\varphi _m (\tau )} \right\rangle  \to e^{if_m (\tau )}
\left| {\varphi _m (\tau )} \right\rangle \;\;(f_m (0) = 0).
\end{equation}
Here $f_m (0) = 0$ is because of given initial state. We call this
adiabatic orbit with special choice of the time-dependent phase
factor as the $U(1)$-$invariant$ $adiabatic$ $orbit$.

It is clear that, although the initial conditions $\left| {m,0}
\right\rangle$ are the same, the dynamic evolution orbit
$|\Phi_m(\tau)\rangle$ do not always coincide with the adiabatic
orbit $|\varphi_m(\tau)\rangle$, or they are not even close to each
other. Obviously they coincide if and only if

\begin{equation}
\gamma _{nm} = 0\;\;(\forall n \ne m). \label{a3}
\end{equation}

Generally speaking, the dynamic evolution orbit $\left| {\Phi _m
\left( \tau  \right)} \right\rangle$ starting from the initial state
$\left| {m,0} \right\rangle $ will change among some adiabatic
orbits which will cause transitions between different orbits. Our
task is to find the proper condition under which the dynamic orbit
is sufficiently close to the adiabatic orbit when the Eq.(\ref{a3})
is not satisfied. Since the Hamiltonian $h(\tau)$ is Hermitian, all
the $U(1)$-$invariant$ $adiabatic$ $orbits$ in Eq.(\ref{a4}) at a
given time constitute a complete orthonormal basis of the system. In
this basis, the dynamic evolution orbit of the system reads
\begin{equation}
\left| {\Phi _m \left( \tau  \right)} \right\rangle  = \sum\limits_n
{c_n (\tau )\left| {\Phi _n^{adia} \left( \tau  \right)}
\right\rangle \;,\;\;\;\;}  {\left| {\Phi _m \left( 0 \right)}
\right\rangle }  = \left| {m,0} \right\rangle. \label{a6}
\end{equation}
The expansion in Eq.(\ref{a6}) is referred to as the
$U(1)$-$invariant$ $adiabatic$ $expansion$ with the time-dependent
coefficients. Therefore, the set of coefficients equations reads

\begin{equation}
\dot c_m (\tau ) = i\sum\limits_{n \ne m} {c_n (\tau )M(\tau )_{mn}
} \;,\label{t7}
\end{equation}
where the diagonal elements of the matrix $M(\tau )$ are zero and
the non-diagonal elements of $M(\tau )$ read
\begin{equation}
M(\tau )_{mn}  = i\left\langle {{\Phi _m^{adi} (\tau )}}
 \mathrel{\left | {\vphantom {{\Phi _m^{adi} (\tau )} {\dot \Phi _n^{adi} (\tau )}}}
 \right. \kern-\nulldelimiterspace}
 {{\dot \Phi _n^{adi} (\tau )}} \right\rangle  \equiv \left| {\gamma _{mn} (\tau )} \right|e^{i\theta _{mn} (\tau
 )}, \label{t8}
 \end{equation}
where
\begin{equation}
\theta _{mn} (\tau ) = \int_0^\tau  {d\lambda \left( {e_m (\lambda )
- e_n (\lambda ) + \gamma _{nn}  - \gamma _{mm} } \right)}  + \arg
\gamma _{mn} (\tau ).
 \end{equation}

From Eq.(\ref{t7}), it is not hard to get
\begin{equation}
\mathord{\buildrel{\lower3pt\hbox{$\scriptscriptstyle\rightharpoonup$}}
\over C} (\tau )^{(n + 1)}  = A(\tau
)\mathord{\buildrel{\lower3pt\hbox{$\scriptscriptstyle\rightharpoonup$}}
\over C} (\tau ) \label{d1}
\end{equation}
where
\begin{equation}
\begin{array}{l}
 A(\tau ) = \sum\limits_{\sum\limits_{h = 1}^k {\left( {j_h  + 1} \right)}  = n + 1} {\frac{{n!i^k }}{{\prod\limits_{p = 1}^k {j_p !} \prod\limits_{l = 1}^{k - 1} {\left( {n + 1 - \sum\limits_{m = 1}^l {(j_m  + 1)} } \right)} }}}  \\
 \;\;\;\;\;\;\;\;\;\;\;\;\;\;\;\;\;\;\;\;\;\; \cdot M^{(j_1 )} M^{(j_2 )}  \cdots M^{(j_k )}  \\
 \end{array}
 \end{equation}
and $(n+1)$ denotes the $(n + 1)^{th}$ derivation of $\vec C (\tau)$
with respect to time. Then we have following theorem

{\bf Theorem} {\it For an $N$-level quantum system and an arbitrary
real $\varepsilon$ and a time period $T$, if the following
conditions hold
\begin{eqnarray}
&\mathop {\max }\limits_{\forall n,i,j} A(0)_{ij} \;and\;\mathop
{\max }\limits_{\forall n,i,j} A(T)_{ij}  < B < \infty\;\;\;\;\;\;\;\;\;\;\;\;\;\;\;\;&\\
&\begin{array}{l}
 \left| {\gamma _{mn} (\tau )} \right|e^{i\dot \theta _{mn} } \;can\;be\;expanded\;as\;\sum\limits_{k = 1}^p {a_{mn}^k e^{i\omega _{mn}^k \tau } }\;\;\;  \\
 with\;\mathop {\max }\limits_{\forall n \ne m,k} \left| {a_{mn}^k } \right| = D < \infty\;  \\
 \end{array}&\\
&\mathop {\max }\limits_k \left| {\frac{1}{{\omega _{mn}^k }}}
\right| \le \frac{\varepsilon }{{2pN(N - 1)BD}} \equiv \varepsilon '
< 1\; with \;\varepsilon  + \varepsilon ' \le 1.&
\end{eqnarray}
then the probability of finding dynamical orbit in the adiabatic
orbit $\left| {\Phi _m^{adi} (\tau )} \right\rangle$ is greater than
$\left( {1 - \delta } \right)^2 \;with\;\delta  = \varepsilon /(1 -
\varepsilon ')$.}

$Proof$: From Eq.(\ref{t7}) and the conditions above, we have
\begin{equation}
\begin{array}{l}
 \left| {c_m (T) - 1} \right| = \left| {\sum\limits_{n \ne m} {\int_0^\infty  {d\tau \left( {c_n (\tau )\sum\limits_{k = 1}^p {a_{mn}^k e^{i\omega _{mn}^k \tau } } } \right)} } } \right| \\
  = \left| {\sum\limits_{n \ne m} {\sum\limits_{k = 1}^p {a_{mn}^k } \sum\limits_{q = 0}^\infty  {( - 1)^q \frac{{c_n (T)^{(q)} e^{i\omega _{mn}^k T}  - c_n (0)^{(q)} }}{{\left( {i\omega _{mn}^k } \right)^{q + 1} }}} } } \right| \\
  \le \sum\limits_{n \ne m} {\sum\limits_{k = 1}^p {\sum\limits_{q = 0}^\infty  {\left| {2NBD\left( {\frac{\varepsilon }{{2pBDN(N - 1)}}} \right)^{q + 1} } \right|} } }  = \frac{\varepsilon }{{1 - \varepsilon '}} \\
 \end{array}
\end{equation}
Namely,
\begin{equation}
1 - \left| {c_m (T)} \right| \le \left| {1 - c_m (T)} \right| \le
\frac{\varepsilon }{{1 - \varepsilon '}}.
\end{equation}
Therefore, the probability of finding dynamical orbit in the
adiabatic orbit $\left| {\Phi _m^{adi} (\tau )} \right\rangle$ is
\begin{equation}
P_m (T) = \left| {c_m (T)} \right|^2  \ge \left( {1 -
\frac{\varepsilon }{{1 - \varepsilon '}}} \right)^2.
\end{equation}
Thus we prove the theorem.

Although Eq.(12-14) in the $theorem$ are sufficient, however, it is
somewhat too complicated. It is not hard to find for any $N$-level
Hamiltonian with both time-independent terms of $| \gamma_{nm}|$ and
$\dot \theta_{nm}$ satisfying Rydberg-Ritz Combination
Principle(RRCP) $\dot \theta _{nl}  + \dot \theta _{lm} = \dot
\theta _{nm}$, for an arbitrary real $0 \le \delta  \ll 1$, when the
following condition holds

\begin{equation}
\max _{\forall m\;and\;k,n \ne m} \frac{{\left| {\gamma _{km} }
\right|}}{{\left| {\dot \theta _{nm} } \right|}} \le \frac{\delta
}{{\sqrt {N - 1} }}\label{m11}
\end{equation}
viz.,
\begin{equation}
\max _{\forall m\;and\;k,n \ne m} \frac{{\left| {\gamma _{km} }
\right|}}{{\left| {e_n (\tau ) - e_m (\tau ) + \Delta _{mn} (\tau )}
\right|}} \le \frac{\delta }{{\sqrt {N - 1} }} \label{q11}
\end{equation}
where
\begin{equation}
\Delta _{mn} \left( \tau  \right) \equiv  {\gamma _{mm} \left( \tau
\right) - \gamma _{nn} \left( \tau  \right) + \frac{d}{{d\tau }}\arg
\gamma _{nm} \left( \tau  \right)}\;\;(\forall n \ne m). \label{c13}
\end{equation}
then the probability of finding dynamical orbit in the adiabatic
orbit $\left| {\Phi _m^{adi} (\tau )} \right\rangle$ is greater than
$(1 - \delta )^2$.

\emph{Proof}: \  Denote $c'_{m}(\tau)=e^{i\omega_m \tau}c_m(\tau)$
with $\omega_m-\omega_n=\dot{\theta}_{mn}$. Then, $c'_m(\tau)$
satisfy equations
\begin{eqnarray}
i\frac{\partial}{\partial \tau}\vec{C'}(\tau)=\Pi \vec{C'}(\tau)
\end{eqnarray}
where $\vec{C'}(\tau)=(c'_1(\tau),c'_2(\tau),\ldots,c'_N(\tau))^T$,
$\Pi$ is a self-adjoint matrix, $\Pi_{kk}=\omega_k$ and
$\Pi_{kl}=|\gamma_{kl}|$. Denote eigenvalues of $\Pi$ as $\eta_m$,
we have \cite{Proof}
\begin{equation}
|\eta_m-\omega_m|\leq\sqrt{\sum\limits_{k\neq m}2|\gamma_{km}|^2}.
\label{final}
\end{equation}

If unitary matrix $U$ diagonalizes $\Pi$, then $U\Pi
U^{\dag}=diag\left\{ {\eta _1 ,\eta _2 , \cdots ,\eta _N }
\right\}$, that is $U_{ik}\Pi_{kj}=\eta_i U_{ij}$, thus
$U_{ij}=\sum\limits_{k\neq
j}\frac{|\gamma_{kj}|}{\eta_i-\omega_j}U_{ik}$. When condition
(\ref{q11}) holds, then

\begin{equation}
\frac{{\left| {\gamma _{kj} } \right|}}{{\eta _i  - \omega _j }} \le
\frac{\delta }{{\sqrt {N - 1} }} + \sqrt {\frac{2}{{N - 1}}}
o(\delta ^2 ),\forall i \ne j
\end{equation}
so $|U_{ij}|  \le \left( {\delta  + \sqrt 2 o(\delta ^2 )}
\right)/\sqrt {N - 1}$ and $|U_{ii}|  \ge 1 - \delta/2  - o(\delta
^2 )$.

 $\vec{C'}(\tau)$ can be solved
exactly as $\Pi$ is time-independent
\begin{equation}
\vec{C'}(\tau)=e^{i\Pi \tau}\vec{C'}(0).
\end{equation}
Applying initial condition $c'_k(0)=\delta_{km}$, the exact solution
of $c'_m(\tau)$ is $\sum\limits_k |U_{mk}|^2 e^{i\eta_k t}$. Thus,
$|c_m(\tau)|\geq 1- \delta$, then the probability of finding
dynamical orbit in the adiabatic orbit $\left| {\Phi _m^{adi} (\tau
)} \right\rangle$ is $P_m (\tau ) \ge (1 - \delta )^2$. Thus the
proof is completed.

The premises of Eq.(\ref{q11}) on Hamiltonian are non-trivial. For
any general 2D system
$h(\tau)=e_+(\tau)|+,\tau\rangle\langle+,\tau|+e_-(\tau)|-,\tau\rangle\langle-,\tau|$,
after applying a transformation $\tau\rightarrow\tau'=g^{-1}(\tau)$
with $g(\tau)\propto\int_0^\tau  {|\left\langle { + ,\lambda }
\right|\frac{d}{{d\lambda }}\left| { - ,\lambda } \right\rangle
|d\lambda }$, then we can forcibly get
$e'_--e'_++\Delta'_{+-}=constant$ which makes the final
time-dependent 2D system satisfying the limitations of
Eq.(\ref{q11}).

In condition Eq.(\ref{q11}), there appears a new interesting
quantity $\Delta _{mn}$ referred to as $quantum$ $geometric$
$potential$ (QGP) for following three reasons. First, QGP is also
$U(1)$-invariant under the transformation Eq.(\ref{gauge}). Second,
 the integral of QGP
over a closed smooth curve is the difference of Berry phases of
different adiabatic orbits. And the last reason is that the value of
QGP depends only on the path and measure of adiabatic orbit or, in
other words, $\Delta _{mn} (\tau )/\left| {\gamma _{mn} }
\right|\;(\forall n \ne m)$ is invariant under any transformation
$\tau\rightarrow \tau'=f(\tau)$. Furthermore, It can be proved that
in 2D systems $\Delta _{mn} (\tau )/2\left| {\gamma _{mn} } \right|$
is just the geodesic curvature of spherical curve corresponding to
the adiabatic orbit on the surface of Bloch sphere or 2D real Ray
space.

\emph{Proof}: Generally, we can write the Hamiltonian of a 2D system
as $h\left( \tau\right) = A\left( \tau \right) + B\left( \tau
\right) \vec{n} \left(\tau \right) \cdot \vec \sigma , $ where
$\vec{n} \left(\tau \right) = \left( {\sin \theta \left( \tau
\right)\cos \varphi \left( \tau \right),\sin \theta \left( \tau
\right)\sin \varphi \left( \tau \right),\cos \theta \left( \tau
\right)} \right)$. Choosing appropriate phases, the Hamiltonian's
instantaneous eigenstates or \emph{adiabatic orbits} read
\begin{eqnarray}
\left\{ \begin{array}{l}
\left| { + ,\tau} \right\rangle  = \cos \frac{{\theta \left( \tau \right)}}{2}\left| 0 \right\rangle  + e^{i\varphi \left( \tau \right)} \sin \frac{{\theta \left( \tau \right)}}{2}\left| 1 \right\rangle  \\
\left| { - ,\tau} \right\rangle  = \sin \frac{{\theta \left( \tau \right)}}{2}\left| 0 \right\rangle  - e^{i\varphi \left( \tau \right)} \cos \frac{{\theta \left( \tau \right)}}{2}\left| 1 \right\rangle  \\
\end{array} \right..
\end{eqnarray}
It's quite clear that polarization vectors of the above two
adiabatic orbits point to $\vec{n} \left(\tau \right)$ and $-\vec{n}
\left(\tau \right)$ at time $\tau$, respectively. Considering the
adiabatic orbit $|+,\tau\rangle$, the QGP of this orbit can be
easily calculated as
 \begin{equation}\Delta _{mn}  = \frac{{\dot \theta
\ddot \phi \sin \theta + 2\dot \theta ^2 \dot \phi \cos \theta  +
\dot \phi ^3 \sin ^2 \theta \cos \theta  - \dot \phi \ddot \theta
\sin \theta }}{{\dot \theta ^2  + \left( {\dot \phi \sin \theta }
\right)^2 }}.
\end{equation}
As a comparison, we will calculate the geodesic curvature of the
spherical curve $\vec{r} \left(\tau \right) = \vec{n} \left(\tau
\right)$.
\begin{eqnarray}
\rho  &=& \left( \vec{r} \times \frac{d\vec{r}}{ds} \right) \cdot
\frac{d^2 \vec{r} }{ds^2 } \nonumber \\
&=& \frac{{\dot \theta \ddot \phi \sin \theta  + 2\dot \theta ^2
\dot \phi \cos \theta  + \dot \phi ^3 \sin ^2 \theta \cos \theta  -
\dot \phi \ddot \theta \sin \theta }}{{\left( {\sqrt {\dot \theta ^2
+ \left( {\dot \phi \sin \theta } \right)^2 } } \right)^3 }},
\end{eqnarray}
where curve element $$ ds = \left| d\vec{r} \right| = \sqrt {\dot
\theta  + \left( {\dot \phi \sin \theta } \right)^2 }
d\tau=2|\gamma_{mn}|d\tau. $$ Then we get
\begin{equation}\frac{\Delta _{mn}}{2|\gamma_{mn}|} = \rho .\end{equation}

In the following part, two models will be presented to show
Eq.(\ref{q11}) is a good sufficient adiabatic condition and the
effect of QGP is significant. Firstly, we shall study a spin-half
particle in a magnetic field. The Hamiltonian of the system is
\begin{equation}
h(\tau ) = \eta \sigma _z  + \xi \left[ {\sigma _x \cos \left(
{2K\eta \tau } \right) + \sigma _y \sin \left( {2K\eta \tau }
\right)} \right],
\end{equation}
where $\eta  = \hbar \omega _0 /E_ \pm  ,\;\xi  = \hbar \omega /E_
\pm \;{\rm{and}}\;E_ \pm   = \sqrt {\eta ^2  + \xi ^2 }
\;{\rm{are}}\;{\rm{all}}\;{\rm{constants}}$. Obviously,
Eq.(\ref{q11}) is a sufficient adiabatic condition for this kind of
Hamiltonian. Properly choosing phases, two adiabatic orbits can be
written as
\begin{eqnarray}
\left\{ \begin{array}{l} \left| { \varphi_+(\tau) } \right\rangle  =
\cos \left( {\frac{\theta }{2}}
 \right)\left| 0 \right\rangle  + e^{2iK\eta \tau } \sin \left( {\frac{\theta }{2}} \right)\left| 1 \right\rangle  \\
\left| { \varphi_-(\tau) } \right\rangle  = \sin \left( {\frac{\theta }{2}} \right)\left| 0 \right\rangle  - e^{2iK\eta \tau } \cos \left( {\frac{\theta }{2}} \right)\left| 1 \right\rangle  \\
\end{array} \right.,
\end{eqnarray}
where $\cos \theta  = \eta /\sqrt {\eta ^2 + \xi ^2 }$. Consider the
adiabatic orbit $|\varphi_+(\tau)\rangle$, we have QGP, $\Delta
_{+-} = 2K\eta \cos \theta $. It is easy to obtain the expression of
the new adiabatic condition of Eq.(\ref{q11})
\begin{equation}
\left| {\sqrt {\eta ^2  + \xi ^2 }  - K\eta \cos \theta } \right|
\gg \left| {K\eta \sin \theta } \right|.
\end{equation}
Suppose the initial state of the system is $|+,0\rangle$, we have
the fidelity between the dynamic evolution orbit and the adiabatic
orbits at time $\tau$
\begin{eqnarray}
\begin{array}{l}
 F(\tau ) = \sqrt {\cos ^2 (A\tau ) + \sin ^2 (A\tau )\left[ {\frac{{(1 - K)\eta \cos \theta  + \xi \sin \theta }}{A}} \right]^2 }  \\
 \end{array}
 \end{eqnarray}
where $A = \sqrt {\left( {1 - K} \right)^2 \eta ^2  + \xi ^2 } $ is
also a constant parameter.

If we choose $\eta\gg\xi$ and $K\simeq1$, then the traditional
condition [16] is satisfied but the new condition Eq.(\ref{q11}) is
not. Meanwhile, the fidelity $F\left( \tau \right) \approx \sqrt {1
- \cos ^2 \theta \sin ^2 \left( {A\tau } \right)} \nrightarrow 1$
when $\tau$ is not too small. Thus, even though the traditional
condition is satisfied and we might regard the system as slowly
changing one, the quantum adiabatic approximation may be unfaithful
description of the system because of the effect of the QGP.

While if we choose $\eta\gg\xi$ with $K\gg1$ and $K\gg \eta$, in
this case, the QGP is much larger than the difference of the
instantaneous energy eigenvalues, and the new condition
Eq.(\ref{q11}) is satisfied while the traditional one is not. Now we
have $F\left( \tau \right) \approx \sqrt {1 - \sin ^2 \theta \sin ^2
\left( {A\tau } \right)} \approx 1$. Therefore, the QGP can help to
guarantee the validity of the adiabatic approximation despite the
difference of energy eigenvalues is too small to satisfy the
traditional condition.

Next, notifying that if QGP has same sign as the corresponding
difference of energy eigenvalue, it will positively guarantee the
system evolution to be adiabatic. Moreover, if $\left| {\Delta _{nm}
} \right|/\left| {\left\langle {n}
 \mathrel{\left | {\vphantom {n {\dot m}}}
 \right. \kern-\nulldelimiterspace}
 {{\dot m}} \right\rangle } \right| \gg 1$, with
Eq.(\ref{q11}), then the evolution may be adiabatic whether the
adiabatic orbit moves slowly or fast. Thus we shall present an
interesting Hamiltonian for illustrating QGP may be helpful to
construct robust system. Consider a 2D system governed by
Hamiltonian $h(\tau)=\eta \sigma_z+e^{-i\eta \sigma_z \tau}(\eta_0
\sigma_x + \eta_1 e^{i\eta_2 \sigma_x \tau}\sigma_ze^{-i\eta_2
\sigma_x \tau})e^{i\eta \sigma_z \tau}$ with $\eta_0/\eta\gg 1 \
\text{and }\eta_0/\eta_1\gg 1$. The density matrixes of adiabatic
orbits read
\begin{equation}
\rho^{adi}_\pm(\tau)=\frac{1}{2}\pm\frac{\eta \sigma_z+e^{-i\eta
\sigma_z \tau}(\eta_0 \sigma_x + \eta_1 e^{i\eta_2 \sigma_x
\tau}\sigma_ze^{-i\eta_2 \sigma_x \tau})e^{i\eta \sigma_z
\tau}}{2N(\tau)},
\end{equation}
where $N(\tau)=\sqrt{\eta_0^2+(\eta+\eta_1\cos2\eta_2
\tau)^2+\eta_1^2\sin^2 2\eta_2 \tau }$. The density matrixes of the
evolution orbits starting from the corresponding initial states of
adiabatic orbits reads
\begin{equation}
\rho_\pm(\tau)=\frac{1}{2}U(\tau)
\left(1\pm\frac{\eta_0\sigma_x+(\eta_1+\eta)\sigma_z}{N(0)}\right)U^{\dag}(\tau),
\end{equation}
where $U(\tau)=e^{-i\eta \sigma_z \tau}e^{i\eta_2 \sigma_x \tau}
e^{-i((\eta_0+\eta_2)\sigma_x+\eta_1\sigma_z)\tau}$ and the initial
states are $\rho^{adi}_\pm(0)$. The probabilities of staying in the
corresponding adiabatic orbits are
\begin{eqnarray}
P_\pm(\tau)&=&\frac{1}{2N(0)N(\tau)}\left(\eta_0^2+\eta_1^2+\eta^2\cos2\eta_2\tau+2\eta\eta_1\cos^2\eta_2\tau\right. \notag \\
&\ &+\frac{2\tilde{\eta}^4}{\bar{\eta}^2}\sin^2\bar{\eta}\tau-
\frac{4\eta(\eta_0+\eta_2)\tilde{\eta}^2}{\bar{\eta}^2}\cos^2\eta_2\tau\sin^2\bar{\eta}\tau \notag\\
&\
&+\left.\frac{\eta\tilde{\eta}^2}{\bar{\eta}}\sin2\bar{\eta}\tau\sin2\eta_2
\tau\right)+\frac{1}{2}.
\end{eqnarray}
Here $\bar{\eta}=\sqrt{\eta^2_1+(\eta_0+\eta_2)^2}\ \text{and} \
\tilde{\eta}=\sqrt{\eta\eta_0+\eta\eta_2+\eta_2\eta_1}$. Since
$\eta_0/\eta\gg 1 \ \text{and }\eta_0/\eta_1\gg 1$, then
probabilities will obtain a lower bound $P_{\min}$ independent on
the magnitude of $\eta_2$: $P_{\min}  = 1 - \left( {\eta  + \eta _1
} \right)^2 /N(0)^2,$ which approach to 1.  It's not hard to verify
when $\eta_2>>\eta$, $\Delta_{+-}$ has the same sign as $E_- - E_+$,
and $\left| {\Delta _{ +  - } } \right|/\left| {\left\langle { + }
 \mathrel{\left | {\vphantom { +  {\dot  - }}}
 \right. \kern-\nulldelimiterspace}
 {{\dot  - }} \right\rangle } \right| \simeq \eta _0 /\eta _1  \gg 1$.
When $\eta_2$ is large, the velocity of the adiabatic orbit has the
same order of magnitude of $\eta_2$, at this time, the adiabatic
orbit fast oscillates around the exact dynamic evolution orbit, but
the evolution of the system still keeps adiabatic. Fig.1 shows
evolution orbit and adiabatic orbit for $\eta _0 /\eta  = \eta _0
/\eta _1  = 20\;{\rm{and}}\;\eta _0 /\eta _2  = 0.2$.

\begin{figure}[h]
\begin{center}
\includegraphics[width=0.32\textwidth]{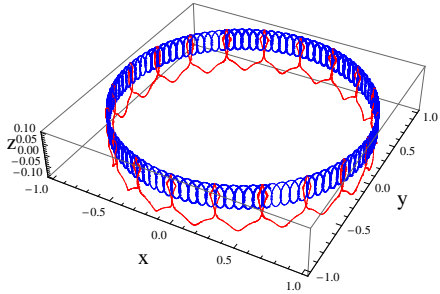}
\end{center}
\caption{evolution orbit(red line) and adiabatic orbit(blue line).}
\end{figure}
This kind of models allows the parameters of the system have a
certain variant range, as long as the adiabatic condition
Eq.(\ref{q11}) holds. Thus we may conclude that the QGP may help
setting up robust systems which may tolerate faults of the system
itself. Another interesting hint from this model is that the
adiabatic orbit may be very complicated comparing with evolution
orbit, which is counterintuitive from the traditional opinion.

For a short summary, it is worthwhile to point that, by the
$theorem$ or new adiabatic condition Eq.(\ref{q11}), the problems
showed in [13,14] has not existed because the relation between
systems $a$ and $b$ constructed in [13,14] does not guarantee them.
Apparently, different from those conditions in \cite{Tong2,
MacKenzie, Vertesi}, our conditions are presented in a more natural
way full of geometric interpretation. One more hint we may get here
is that we should more carefully deal with the phase appearing in
the time-dependent evolution. It is just improperly handling the
phase of Eq.(\ref{t8}) in the work of predecessors \cite{Schiff,
Messiah} that led to their improper traditional condition and later
contradiction in \cite{Marzlin, Tong1}. The condition (\ref{q11})
also implies a modification of the difference of energy eigenvalues
is necessary. Description of the time-dependent evolution might be
more precise and more appropriate via replacing $e_m \left( \tau
\right) - e_n \left( \tau \right)$ by $e_m \left( \tau \right) - e_n
\left( \tau \right) + \Delta_{mn}$. And a related experiment
\cite{Du} for verifying the effect of QGP has been finished. The
experiment also found the characteristic frequency of a kind of
time-dependent systems should be corrected via QGP. The experiment
also illustrated the QGP should reflect some properties of
time-dependent systems, and is not just a convenient mathematical
technique. As it is shown in our paper, QGP may play an important
role in some kinds of time-dependent procedure, but what role it may
play in general time-dependent system is not clear now. We guess
non-trivial QGP will more or less affect the evolution procedure of
time-dependent system.

In conclusion, according to the concepts of $U(1)$-invariant
adiabatic orbit and $U(1)$ invariant expansion stated in this paper,
we present a theorem and a new sufficient adiabatic condition, from
which we get an interesting quantity QGP with its effects and
geometric properties detailedly discussed. At the end we present two
models to show the significant effect of QGP on the evolution.

\acknowledgements

We thank Professor Sixia Yu and Dr. Dong Yang for illuminating
discussions and thank Professor Qimiao Si for suggestions on the
context of this paper. This work is supported by the NNSF of China,
the CAS, and the National Fundamental Research Program (under Grant
No. 2006CB921900).



\begin{thebibliography}{99}
\bibitem{Proof}
Denote $\Pi'=\Pi-B$, where
$\{B\}_{ij}=\Pi_{ij}(\delta_{im}+\delta_{jm}-2\delta_{im}\delta_{jm})$.
Denote $\Pi'$'s eigenvalues as $\nu_i$. One can immediately see
$\nu_m=\omega_m$. Applying the Wielandt-Hoffman theorem on
eigenvalues of self-adjoint matrixes, one
have$\left(\eta_m-\omega_m\right)^2\leq\sum\limits_i\left(\eta_i-\nu_i\right)^2\leq\sum\limits_{k\neq
m}2|\gamma_{km}|^2.$ This is just Eq.(\ref{final}).
\end{thebibliography}
\end{document}